\documentclass[a4paper,12pt]{article}

\usepackage{bm}
\usepackage{mathptmx}       
\usepackage{helvet}         
\usepackage{courier}        
\usepackage{makeidx}         
\usepackage{graphicx}
\usepackage{multirow}        
\usepackage{amsfonts}
\usepackage{amsmath, amsthm, amssymb}
\usepackage{natbib}
\usepackage{color}
\usepackage{bm}
\usepackage{fullpage}
\usepackage{rotating}
\usepackage[english]{babel}
\usepackage[font=small,it]{caption}  
\usepackage{setspace}

\newcommand{\vy}{{\mathbf y}}
\newcommand{\vs}{{\mathbf s}}
\newcommand{\verr}{{\mathbf \epsilon}}
\newcommand{\vbeta}{{\mathbf \beta}}
\newcommand{\vu}{{\mathbf \nu}}
\newcommand{\mx}{{\mathbf X}}
\newcommand{\mw}{{\mathbf W}}
\newcommand{\mi}{{\mathbf I}}
\newcommand{\rY}{{\mathbf Y}}

\newcommand{\comment}[1]{{\color{blue} #1}}
\renewcommand{\comment}[1]{}   

\newcommand{\timeSAOMs}{2 }

            \newcommand{\timeSAOMl}{38 }

\newcommand{\timeSAOMsSD}{0 }

            \newcommand{\timeSAOMlSD}{12 }

            \newcommand{\timeGIBBSssec}{50 }

            \newcommand{\timeGIBBSlsec}{78 }

            \newcommand{\timeGIBBSsSDsec}{11 }

            \newcommand{\timeGIBBSlSDsec}{20 }

\begin{document}

\comment{\tableofcontents}

\title{Modelling Diffusion through Statistical Network Analysis: A Simulation Study} 

\author{Johan A. Elkink\footnote{Corresponding author.}\\University College Dublin\\jos.elkink@ucd.ie \and Thomas Grund\\University College Dublin\\thomas.grund@ucd.ie}

\maketitle


\begin{abstract}

The study of international relations by definition deals with interdependencies among countries. One form of interdependence between countries is the diffusion of country-level features, such as policies, political regimes, or conflict. In these studies, the outcome variable tends to be categorical, and the primary concern is the clustering of the outcome variable among connected countries. Statistically, such clustering is studied with spatial econometric models. This paper instead proposes the use of a statistical network approach to model diffusion with a binary outcome variable. Using statistical network instead of spatial econometric models allows for a more natural specification of the diffusion process, assuming autocorrelation in the outcomes rather than the corresponding latent variable, and it simplifies the inclusion of temporal dynamics, higher level interdependencies and interactions between network ties and country-level features. In our simulations, the performance of the Stochastic Actor-Oriented Model (SAOM) estimator is evaluated. Our simulation results show that spatial parameters and coefficients on additional covariates in a static binary spatial autoregressive model are accurately recovered when using SAOM, albeit on a different scale. To demonstrate the use of this model, the paper applies the model to the international diffusion of same-sex marriage.

\end{abstract}

\newpage

\section{Introduction}

While there is a longstanding literature on policy diffusion and interdependence in political science, international relations, and public policy research, there is a more recent investigation into the statistical methodology appropriate for the study of these effects \citep{BeckGleditschBeardsley2005, FranzeseHays2007, FranzeseHays2008, WardGleditsch2008}, building on the literature on spatial econometrics after the seminal works by \citet{PaelinckKlaassen1979}, \citet{Doreian1980, Doreian1981, Doreian1982}, and \citet{Anselin1988}. In these regression models, the geographical interdependence between units is explicitly modelled and parameters for the level of spatial interdependence estimated. While the original models assume continuous dependent variables, in policy diffusion we often estimate the interdependence in the presence of a policy in some units and its absence in others. In other words, the dependent variable of these models is binary. The same situation occurs when the diffusion of particular regimes is investigated, for example the international diffusion of democracy, or of war and peace \citep{Gleditsch2002, Elkink2011}. To investigate these empirical patterns, standard binomial regression models have been extended for spatial interdependence, starting with \citet{McMillen1992}. An overview of these estimators can be found in \citet{Fleming2004}.

The literature on the statistical analysis of spatial data contains three different approaches to defining geographic space---somewhat akin to how time-series analysis can be modelled in discrete or in continuous time. Data points can be located in continuous space, based on $(x,y)$ coordinates, and thus the statistical models assume the data are a point pattern observed in continuous space. Alternatively, this continuous space can be divided into a regular grid of squares, thus discretizing the data. The third possibility is measuring the geographical proximity or adjacency between each observation \citep{Anselin2002} and study the connection matrix between units. The latter is the typical specification in spatial econometrics---the former two more typically referred to as spatial statistics. The matrix specification, whereby the key geographical nature is captured in an adjacency or distance matrix, can straightforwardly be applied to non-geographical ``distances'', for example capturing trade relations or cultural (dis)similarity measures. It is in this sense that there is a very close conceptual connection between ``space'' and ``network'' in the spatial econometrics literature---the most common specification of space in spatial econometrics is fundamentally that of a network \citep[cf.][fn. 3]{DoreianTeuterWang1984}.

Meanwhile, the literature on statistical network analysis is expanding rapidly, due to advances in big data analysis and the wide range of network data available through social media companies, such as digital friendship ties between users of social media or buying pattern similarities between customers of online stores. This methodology, where complex interdependencies between observations are explicitly taken into account in the model specification, is now also making its impact on political science. For example, the literature starting with \citet{CranmerDesmarais2011} develops the usage of exponential random graph models (ERGMs) into the field of political science. In these models, higher order interdependencies can be explicitly modeled.

The main difference between network statistics and spatial econometrics is that in spatial econometrics, features of the node are the outcome variable, while the network structure is taken as exogenous and is part of the explanation, while in network statistical analysis, the node features are generally taken as exogenous, and the outcome that is explained is the network structure.\footnote{In this paper we treat this as a relatively black-and-white distinction. However, particularly in earlier work, the lines between the two fields blur. Original work in spatial econometrics referring to spatial effects and spatial disturbances models \citep{Doreian1980, Doreian1981, Doreian1982} led to a generalisation called network autocorrelation. In this literature on network autocorrelation, a distinction was made between the network effects model and the network disturbances model (\citealt[156--157]{DoreianTeuterWang1984}; \citealt{Leenders2002}; \citealt{LaRoccaPorzioVitaleDoreian2018}), which became later more commonly known as the spatial autoregressive model and the spatial error model, respectively, in the standard works in spatial econometrics by \citet{Anselin1988, Anselin2002}. Furthermore, extensions to spatial econometric models exist that incorporate endogenous network effects \citep[e.g.][]{FernandezVazquez_etal2009, FranzeseHaysKachi2012, Kachi2012} and to statistical network models that incorporate diffusion effects \citep[e.g.][]{Mohrenberg2013}.} For example, a spatial econometrician might wonder whether neighbouring countries have similar tax regimes \citep{BaturoGray2009}, while a network statistician might investigate whether structural features of the network affects future network tie formation, such as reciprocity effects in political sanctions \citep{CranmerHeinrichDesmarais2014}.

In a diffusion application of a spatial econometric model we typically have observations at fixed time periods. For example, we might have annual data on whether a particular political regime is a democracy or not---typical in a study of the diffusion of democracy. The speed at which the diffusion takes place might be much faster or much slower than annual steps, however. A revolution in one country can lead to protests and a collapse of an autocratic regime in a neighbouring country within weeks or months. In the data set, however, we do not observe any changes or dynamics between the annual time points---we only observe the new state at the end of the year. This discretization of time renders it logically impossible to model autocorrelation in the binary outcome variable. Instead, spatial econometric models for binary dependent variables model the autocorrelation in the latent, unobserved continuous variable that generalized linear models assume to be underlying the binary observations. The implied assumption is that countries are affected by a neighbour's {\it utility} of a particular potential policy outcome, as opposed to the actual {\it adoption} of that policy. This is implausible in many contexts.

Recent developments in statistical network analysis allow for increasingly complicated patterns, including co-evolution, whereby one type of network tie (e.g. trade) formation is affected by that of another type of network ties (e.g. conflict) and vice versa, or complicated interactions between node- and tie-level variables. Within the social sciences, these dynamics are typically theoretically motivated by assumptions about human interaction and its interdependencies---in other words, it is the node that is the actor, not the tie or the network. This, and the increasing demands on complexity of the models, has lead to the development of the stochastic actor-oriented model (SAOM) \citep{Snijders1996, Snijders2001}. \cite{SnijdersSteglich2015} have also demonstrated how SAOM can be used in cross-sectional settings.

The incorporation of the ability to model both endogenous node-level features and endogenous network tie features simultaneously, in particular through the application of SAOM, leads to the following three key conjectures of this paper:
\begin{enumerate}
	\item Statistical network models (in particular, SAOM) should be able to capture and estimate spatial interdependencies such as those typically modelled using spatial regression models (in particular, spatial autoregression with binary outcomes).
	\item If we can use statistical network models to investigate diffusion of policies, regimes, or conflict, we will have a better tool available to investigate higher level interdependencies, temporal dynamics, and network co-evolution in the study of diffusion.
	\item The diffusion of policies, regimes, or conflict can be more plausibly modelled as autocorrelation in the actual outcomes, as opposed to the underlying utilities, when using statistical network models.
\end{enumerate}
Because of the logic of the SAOM model, where actors evaluate distinct proposals for change, there is a restriction on the node-level features that only allows for binary or categorical variables as the outcome variable.\footnote{Recent developments attempt to relax this restriction.} Thus, while the binary nature of the outcome variable in spatial autoregressive models generates particular challenges for estimation \citep{Fleming2004, Bille2013, CalabreseElkink2014}, in a SAOM specification, binary outcomes at the node-level are the default.

\citet{Mohrenberg2013} provides a further argument for the use of network models to study processes of diffusion, namely the ability to control for selection mechanisms when estimating influence patterns---are neighbours more similar because they adapt to each other, or are they neighbours because they are similar? When geographic ties are used, this argument is irrelevant, but when for example trade or diplomatic ties are used to explain policy diffusion or regime similarity, as in \citet{Mohrenberg2013}, then this is an important additional reason to use network rather than spatial model specifications. While the fact that both can be incorporated in the same model is indeed a key reason to prefer statistical network models over spatial econometric specifications,\footnote{See \citet{Kachi2012} for an extension of spatial econometrics exactly to address this concern, however.} this paper focuses on the comparison in estimation quality assuming the spatial autoregressive (SAR) context---i.e. a fixed, exogenous network and a static, cross-sectional estimation of spatial or network autocorrelation.

The primary analysis in this paper is therefore a Monte Carlo simulation analysis of data generated with the  binary spatial autoregressive model in mind, with the spatial effect estimated using the SAOM model specification. \comment{Another possibility for further research could be to simulate data with the SAOM and see if we can recover the effects with the spatial model. Just flip it around.} The paper proceeds with first an introduction to binary spatial autoregressive (BSAR) models, statistical network models, and a comparison of the BSAR and SAOM models. It then provides a simulation study of using SAOM models on data generated using the BSAR data generation process, demonstrating that the autoregressive parameters are properly recovered---albeit on a different scale---and that coefficients on other covariates in the model are accurately estimated. This suggests that the SAOM model can be used for BSAR contexts, even in cross-sectional analyses, while providing a number of additional benefits such as the ability to model co-evolution and naturally extend the model to incorporate time-series dynamics. Most importantly, it implies a more natural way to model diffusion of binary outcomes, whereby neighbouring countries react to outcomes rather than utilities in a country.

\section{Binary spatial autoregressive models}

Spatial econometric models---models whereby spatial or network interdependence is explicitly modelled in a regression specification---are typically specified as \[ \rY=\mx\vbeta+\rho\mw\rY+\vu, \] with \[ \vu=\gamma\mw\vu+\verr, \] where $\mx$ is an $n\times k$ design matrix and $\verr \sim N(0,\sigma_{\verr}^2)$. Here, $\mw$ is an exogenously given contiguity matrix with $w_{ij}\neq 0$ if units $i$ and $j$ are adjacent or proximate and $w_{ij}=0$ when they are not ($w_{ii}=0$ for all $i$). Often a normalized matrix is used such that $\sum_j w_{ij}=1$ for all $i$ \citep{Tiefelsdorf2000}---for a critical review of this practice, see \citet{NeumayerPluemper2016}.\footnote{Although much less common, one could also normalize over the columns, such that $\sum_i w_{ij}=1$. In the first scenario, influence decreases as more actors influence $i$, in the second scenario, influence decreases as $j$ influences more actors \citep[34]{Leenders2002}.} With this normalization, $\mw\vy$ amounts to calculating, for each observation, the average value of neighbouring regions on the dependent variable.\footnote{Throughout, we use uppercase letters to refer to matrices and random variables, lowercase to refer to observed or specific values.} With this generic specification, $\gamma=0$ results in the spatial autoregressive (SAR) model and $\rho=0$ in the spatial error model (SEM). This paper is solely concerned with the SAR specification. Because the influence of neighbours operates in both directions and simultaneously, this SAR model specification leads to heteroskedastic errors: \begin{equation} \rY=(\mi-\rho\mw)^{-1}\mx\vbeta + (\mi-\rho\mw)^{-1}\verr, \label{eq:hetero}\end{equation} which has to be accounted for in the estimation procedure. In the spatial econometric literature, the standard references are \citet{Doreian1980, Doreian1981, Doreian1982} and \citet{Anselin1988}. The same model is well known in the social network literature, where it is more appropriately known as the network autocorrelation model \citep{DoreianTeuterWang1984, Leenders2002, LaRoccaPorzioVitaleDoreian2018}.

When the dependent variable is binary, one can use a binary spatial autoregressive (BSAR) model, with \begin{equation} \rY^*=\mx\vbeta+\rho\mw\rY^*+\verr \label{eq:bsar}\end{equation} and observation mechanism \[ Y_i = \left\{
           \begin{array}{ll}
             1 & \text{if}\quad Y_i^*>0 \\
             0 & \text{otherwise.}
           \end{array}
         \right. \] $\rY^*$ is here a continuous random vector that could be interpreted as the utility of a positive outcome. The implied heteroskedasticity leads to greater complications than for the continuous case \citep{McMillen1992, McMillen1995}. A multivariate normal distribution for $\verr$ here results in a spatial probit model and a multivariate logistic distribution in a spatial logit. \citet{Fleming2004} provides an overview of the various estimators proposed for the BSAR model, while \citet{Bille2013} and \citet{CalabreseElkink2014} perform Monte Carlo simulation studies to evaluate their respective ability to estimate the intensity of the spatial interdependence. While various models have been proposed that perform reasonably well, these estimators tend to be computationally intensive. An exception is the linearized general method of moments estimator proposed by \citet{KlierMcMillen2008}, which is fast and performs well as long as the sample size is large and the intensity of the spatial autocorrelation relatively low \citep{CalabreseElkink2014}. In the social network literature, \citet{Zhang_etal2013} discuss the network autocorrelation model with binary outcomes, which basically re-introduces the expectation-maximization algorithm by \citet{McMillen1992, McMillen1995} and the Gibbs estimator by \citet{LeSage2000} and \citet{SmithLeSage2004}, but for models with multiple connectivity matrices \citep[cf.][]{FranzeseHaysKachi2012, Kachi2012}.\footnote{The {\tt tnam} package in R \citep{LeifeldCranmer2017} is implemented for both continuous and binary dependent variables, but in the case of contemporaneous spatial effects, where the time lag is zero, this model includes an exogenous $\mw\vy$ as a covariate to the model, without addressing the implied endogeneity and resulting heteroskedasticity outlined in the literature on network autocorrelation and spatial regression models.}

An important observation is that on the right-hand side we use $\mw\rY^*$ rather than $\mw\rY$, so it is not the proportion of neighbours that, for example, adopted a particular policy, but the average latent variable among neighbours that creates the contagious pressure. In the spatial econometric context the alternative is ``infeasible'' \citep[171, fn. 1]{BeronVijverberg2004} and ``algebraically inconsistent'' \citep[8]{KlierMcMillen2008} due to the discretization of time. While $Y_{it}$ is a function of $Y^*_{it}$, $Y^*_{it}$ would simultaneously  be a function of $Y_{it}$. Only if one precedes the other would it be possible, as $Y^*_{it}$ would be a function of $Y_{i,t-1}$ instead. Imagine Germany makes a transition to a flat tax policy in February and the Netherlands follows the example by making a transition in October of the same year. In practice this could be a genuine reaction to the Germany policy change, but in the model, due to the discretisation of time, both would take place in the same year, at the same $t$. With temporal autocorrelation, it is often reasonable to assume that it is the underlying utility that is autocorrelated---if a country is likely to adopt a policy at time $t-1$, even if it does not, it might still be likely to adopt that policy at time $t$. In a spatial econometric context, however, it is external countries or actors that react, who might not be able to observe this internal utility or likelihood, and who would more plausibly react to actual outcomes. If one intends to model explicitly the copying of actual behaviour, one would have to force a sequence on these events and thus either collect data at much shorter time intervals---which is often infeasible---or model in continuous time or much smaller time steps, so that one can use a lagged dependent variable---which is the approach we propose here.

Although there is typically a dynamic interaction implied in the theory, related to learning, competition, or other type of reaction by neighbours, these model specifications assume a cross-sectional observation of the data. In the analysis below we similarly assume static data, since the purpose is to evaluate the extent to which statistical network estimators can estimate the same model as what the spatial econometric models are designed to estimate, but the extension to a dynamic time-series analysis is significantly more straightforward in SAOM than in the BSAR model.\footnote{See \citet{FranzeseHaysCook2016}, however, for an extension of the BSAR specification to a panel data context, in discrete time, using the best performing estimator in \citet{CalabreseElkink2014}, namely the recursive importance sampler developed by \citet{BeronVijverberg2004}.}

\section{Statistical network models}


While a range of variations are available, the main statistical models used for the prediction of tie formation based on structural features of the network are the exponential random graph model (ERGM) and the stochastic actor-oriented model (SAOM). Both have important features in common, in that they are designed for the prediction of binary ties between nodes, using network statistics such as reciprocity, similarity, transitivity, and many other \citep{LeifeldCranmer2016}. For the ERGM model, this amounts to a probability density function \[ p(\mw, \vbeta)=\frac{\exp\{\vs(\mw)\vbeta\}}{\sum_{\mw^*\in \Psi} \exp\{\vs(\mw^*)\vbeta)\}}, \] where vector $\vs(\mw)$ captures a range of statistics on network $\mw$ and $\Psi$ is the set of all possible networks. The problem of having a high level of interdependence between observations is therefore solved by taking the network itself as the unit of analysis and all possible networks as the sample space, as opposed to all possible values for a particular tie.

In a SAOM model, the idea is that actors are in control of their network ties and behaviours given the structural constraints \citep[see][]{Udehn2002, Hedstrom2005} and that the observed network is the outcome of actor-level choices. Such an actor-oriented approach is reflected in the way in which actors are given opportunities to change their network ties, i.e. formation of a new tie or dissolution of an existing tie, or their behaviour, i.e. increment and decrement a behaviour value.\footnote{In SAOM models the term 'behaviour' is used to refer to all sorts of actor-level attributes that can change.}
\comment{This ability to model behaviour at the node-level, rather than inferring node-level behaviour from aggregative level statistics such as common in the ERGM model---an inference that is problematic in a similar vein to problems of ecological inference \citep[cf.][]{Block_etal2017}---constitutes  the main difference between ERGM and SAOM.} Both the formation or dissolution of a tie and changes in actor attributes are modelled as a Markov process through so-called \textit{ministeps} \citep[see][]{HollandLeinhardt1977, Wasserman1979}, where only one change occurs at a time. Thus, only the current state of the network and actor attributes is assumed to probabilistically determine its further evolution.\footnote{For a more detailed discussion of these assumptions, see \citet{Snijders_etal2010}.} Historically, the ERGM model was used for cross-sectional and the SAOM for longitudinal data, but various extensions of the ERGM model \citep[see, e.g.,][]{Hanneke_etal2010, CranmerDesmarais2011, Koskinen_etal2015} as well as applications of the SAOM model to cross-sectional data, as we will show, makes this distinction obsolete.\footnote{Like space, time can be modelled as continuous time or in discrete time steps. In statistical network analysis this distinction is visible in the separation between TERGM \citep{CranmerDesmarais2011} and SAOM models, whereby the latter assumes continuous time \citep{Block_etal2018}.}

The SAOM model was introduced by \citet{Snijders1996} to study the evolution of networks from one network panel wave to the next and the co-evolution of network dynamics and actor attributes \citep[see also][]{Snijders2001, Snijders_etal2010}. Changes in the network structure can be due to the structural position of the actors within the network at a previous point in time, for example, when somebody becomes the friend of a friend. They can also be due to characteristics of the actors, for example, actors with certain attributes might tend to create more new ties. Lastly, network change might be driven by the characteristics of pairs of actors. Similarly, changes in node-level behaviours can be the outcome of the structural position of the actors, such as popular nodes being more likely to show particular behaviours. Social influence is another example where behaviour is affected by network position. The ability to model network dynamics and behaviour dynamics simultaneously makes the SAOM model also attractive for the separation of social selection and social influence effects \citep{Steglich_etal2010, Mohrenberg2013}.

The SAOM model can be understood as an agent-based simulation model (ABM) \citep{MacyWiller2002}, which only deviates from traditional ABMs in the way that it is used for statistical inference. The agent-based model thought to generate the network and node-level behaviours includes two sub-processes. In the first process, actors are sampled and given the opportunity to make changes to network ties or behaviours that they control. The opportunity to make changes between  observed panel waves are modelled by the so-called \textit{rate function}. Again, the network position of an actor or other attributes may play a role for making actors more or less likely to make changes between panel waves. In SAOM models which include both network and actor attribute dynamics, so-called co-evolution models, there exists a \textit{network rate} and a \textit{behaviour rate function}. Effectively, these two rate functions repeatedly sample one focal actor $i$ in continuous time, who is given the opportunity to make either one network or one behaviour change, respectively.

After a focal actor $i$ has been given the opportunity to make a change, the second process describes the tie or behaviour change the actor makes. In the case of a network change, a \textit{ministep} could be 1) creating a new tie to somebody the focal actor is not already connected with; 2)  deleting one of the focal actor's outgoing ties, which the focal actor is assumed to be in control of; or 3) maintaining the status quo. Following the notation above, we denote such a proposed network state $\mw'$, which only differs from the current network state $\mw$ by one \textit{ministep} the actor $i$ can potentially execute. In the case of a behaviour (or attribute) change, such a \textit{ministep} would be either 1) incrementing the attribute variable; 2) decrementing the attribute variable; or 3) maintaining the status quo.\footnote{In the case of binary variables, the options are limited accordingly to reflect the boundaries of possible variable values.} We denote the proposed attributes $\vy'$, in contrast to the current state of the actor attributes $\vy$. Again, $\vy'$ only differs from $\vy$ by one \textit{ministep} executed by actor $i$, hence, actor $i$ can only propose changes to actor $i$'s attribute.

Each change that is executed by an actor alters the state of the network as a whole, which then impacts the conditions in which the next actor can make a change. Hence, the current states $\mw$ and $\vy$ are continuously updated after each \textit{ministep} and only then the next actor is sampled to potentially make the next change.

All possible changes that a focal actor can make are evaluated and assigned an objective score, which reflects the attractiveness of a certain change compared to others. These scores are derived through the  \textit{objective function}, which includes change statistics and parameters, which are to be estimated. Again, in co-evolution models, there is a \textit{network objective} and a \textit{behaviour objective function}, describing the desirability of the proposed changes $\mw'$ and $\vy'$ respectively.\footnote{Further extensions of the model allow to distinguish between creation and dissolution of ties. In this case there is a \textit{creation function} and a \textit{maintenance} (or \textit{endowment}) \textit{function}, which are analogous to the network objective function, but only give scores for tie creation or deletion, respectively.} Formally, the objective functions are written as the linear combinations:

 \[ {f_i^{net}}(\mw',\vy,\vbeta^{net}) =  \sum_{k}\beta_k^{net}s_{ki}^{net}(\mw', \vy) \]

 \[ {f_i^{beh}}(\mw,\vy',\vbeta^{beh}) =  \sum_{k}\beta_k^{beh}s_{ki}^{beh}(\mw, \vy') \] where, $f_i^{net}$ and $f_i^{beh}$  are the values of the objective function (or the attractiveness score) the focal actor $i$ attributes to each potential new state of the network or behaviour change, respectively. The functions $s_{ki}(\mw, \rY)$ are effects that are included in the model and are similar to independent variables in a normal regression. For example, a new tie might close two triads, hence, the underlying statistic for the number of triads in the network would increases by two because of this change. Once all options are assigned an objective score, the actual choice is made using a \citet{McFadden1973} random utility formulation. For an actor $i$ who has been given the opportunity to make a change in a \textit{ministep}, which would transform the network into one of $\Phi$ possible new states, the probability for making a specific change towards the network state $\mw'$ is given by: \[ p_i^{net}(\mw', \vbeta^{net})=\frac{{f_i^{net}}(\mw',\vy,\vbeta^{net})}{\sum_{\mw^*\in \Phi} {f_i^{net}}(\mw^*,\vy,\vbeta^{net})}, \] Similarly, the probability that actor $i$, when given the opportunity to make a behavioural change, transforms its behaviour into one of $\Omega$ possible states (whereas each state differs from the current one by one behavioural \textit{ministep}) is given by: \[ p_i^{beh}(\vy', \vbeta^{beh})=\frac{{f_i^{beh}}(\mw, \vy',\vbeta^{beh})}{\sum_{\vy^*\in \Omega} {f_i^{beh}}(\mw,\vy^*,\vbeta^{beh})}, \]

Parameters of these two functions are estimated using conditional multinomial logistic regressions, conditional on the current state of the network. Hence, the estimated parameters for the micro-mechanisms can be interpreted straightforwardly.\footnote{For a full list of effects implemented in the SIENA software package, see \citet{Ripley_etal2018}.} As the choices being made are based on conditional multinomial logistic regressions, the estimated parameters for the micro-mechanisms or tendencies assumed to drive the evolution can be interpreted similar to unstandardised logistic regression coefficients.

\comment{An alternative to the model specification proposed below, which uses SAOM to estimate a spatial autoregressive model, is the autologistic actor attribute model \citep{DaraganovaRobins2013}. This model is based on a discrete rather than a continuous formulation of time, but also on aggregate rather than individual level statistics for modelling behaviour.}

\section{Estimating BSAR using SAOM}

Of the many statistics that can be computed on a network and included as an explanatory variable in a SAOM model, there are two in particular that emulate the spatial autoregressive model. \citet{Mohrenberg2013} suggests the use of the ``average alter effect'' ({\it avAlt}) measure: \begin{equation} s_i^{avAlt}(\mw,\vy)=y_i\cdot \frac{\sum_jw_{ij}y_j}{\sum_jw_{ij}}. \label{eq:avalt} \end{equation} Since both the values in $\mw$ and in $\vy$ can only be zero or one, this amounts to the proportion of neighbouring nodes that have adopted a certain behaviour, for example, having implemented a particular policy, calculated only for those nodes where the policy is implemented.\footnote{Note that the SAOM estimation proceeds by comparing different states of the network, and thus it will compare the state $y_i=1$ to $y_i=0$. The premultiplication with $y_i$ in Eq. \ref{eq:avalt} ensures that $s_i^{avAlt}$ measures the impact of the proportion of neighbours where $y=1$ on switching to or preserving $y_i=1$.} This can be interpreted as a local force to be more or less similar than neighbours. In matrix algebra terms, this would amount to inserting $\mw\rY$ as an explanatory variable in the model, which thus comes very close to, but is crucially different from, the use of $\mw\rY^*$ in Eq. (\ref{eq:bsar}).

An alternative option is the ``average similarity effect'' ({\it avSim}) measure \citep[178]{Ripley_etal2018}: \[ s_i^{avSim}(\mw,\vy)=\frac{\sum_jw_{ij}(\delta^y_{ij}-\widehat{\delta^y})}{\sum_jw_{ij}}, \] where $s_i^{avSim}=0$ if $\sum_jw_{ij}=0$, with \[ \delta^y_{ij}=\delta(y_i,y_j)=1-|y_i-y_j| \] a similarity measure of $y_i$ and $y_j$ and $\widehat{\delta^y}$ the mean similarity score across the network. Since $\delta_{ij}^y$ is one when the two nodes have, for example, the same policy and zero when they do not, this statistic captures the extent to which a node is more dissimilar to neighbours than is the case on average. This can therefore be interpreted as a force to be more or less in line with the global level of clustering. Both measures create a similar effect, but whereas the {\it avSim} drives towards an average level of clustering, thus a high coefficient on this statistic ensures that nodes only become as similar to their neighbours as nodes typically are, the {\it avAlt} drives only in a positive direction, towards more similarity, if the coefficient on this statistic is positive (and more dissimilarity if it is negative).

In addition, regular node-level covariates are entered using the ``main covariate effect'' ({\it effFrom}) measure, which is simply: \[ s^{effFrom}_i=y_ix_i. \] Note the multiplication with $y_i$, similar to the {\it avAlt} specification. Since in each simulation step, one node is randomly identified to make a move, and one proposed move can be a change in $y_i$, the statistic is used to evaluate a hypothetical scenario based on proposal $y_i'$ and a current value $y_i$, so we compare $s'^{effFrom}_i=y'_ix_i$ to $s^{effFrom}_i=y_ix_i$ and thus the multiplication by $y_i$ ensures that the coefficient on this statistic reflects the extent to which $x_i$ contributes to a positive value on $y_i$.

An important difference between the estimation through SAOM as discussed here compared to the BSAR specification in Eq. (\ref{eq:bsar}) is that in both the {\it avSim} and the {\it avAlt} implementations, the average level of similarity or the average value of the neighbour is taken into account on the binary dependent variable, as opposed to the latent variable $\rY^*$. So for {\it avAlt} we have something akin to \[ \rY^*=\mx\vbeta+\rho\mw\rY+\verr \] instead of the specification in Eq. (\ref{eq:bsar}). The infeasibility outlined in the section on the BSAR model is addressed by the use of \textit{ministeps} in SAOM. Instead of estimating the effect of $\mw\rY^*_t$, the average score in neighbouring nodes on the latent variable, we now investigate the effect of $\mw\rY_{\tau-1}$, the proportion of neighbouring nodes that have implemented the policy in the previous \textit{ministep}. A sequence of events is thus explicitly modelled.\footnote{In our Monte Carlo simulations below, since the data generation process makes use of Eq. (\ref{eq:hetero}), this does imply that we would not expect estimates of the spatial autoregressive parameter to be identical to $\rho$ on average, but simply to correlate with it.} This is a major advantage of using SAOM instead of any of the other estimators in \citet{Fleming2004}.

In this paper we are interested only in cross-sectional data---longitudinal extensions of the BSAR model are not yet in common usage---but the continuous time specification does affect the internal dynamics of the SAOM estimator. In order to be able to use the continuous-time SAOM estimator for cross-sectional data, we follow \cite{SnijdersSteglich2015} and implement two ``fake'' time periods, using the same values for $\vy$ and $\mw$ in $t=0$ and $t=1$.\footnote{This implementation requires a small further tweak to the model in order to trick the algorithm into not stopping prematurely. We arbitrarily fix two observations in $\vy$, using mirror images for $t=0$ and $t=1$, such that $y_{1,t=0}=0$, $y_{1,t=1}=1$, $y_{2,t=0}=1$ and $y_{2,t=1}=0$ regardless of the actual observed data. Furthermore, we remove the linear shape effect because there are no real times 0 and 1.} Since we are interested in replicating the spatial configuration, we treat all network ties as structural ties \citep[32]{Ripley_etal2018} in the SAOM model specification. Such structural ties are either structural zeros or structural ones, meaning they will be fixed at these values regardless of any \textit{ministeps}. Since we fix all ties of the network in both $t=0$ and $t=1$, this equates to fixing the \textit{network rate function} at zero. Basically, actors never get the opportunity to change any network ties. At the same time, we fix the \textit{behaviour rate function} at an arbitrary large value to ensure that there will be sufficient \textit{ministeps}.\footnote{In our simulations, we fix the rate such that on average, each node gets one opportunity to change. Different values for this rate lead do not alter our simulation results.} Through this procedure, we obtain a situation where the estimator ``thinks'' there is a time-series, since SAOM is developed for a dynamic context, while in fact it is estimating a cross-sectional model with changes only in node attributes, not the network structure. Furthermore, the fixing of the \textit{behaviour rate function} guarantees that a sufficient amount of changes in the behaviour variable occur before the algorithm tries to return to the original configuration of $\vy$.

\section{Monte Carlo comparison}

The primary purpose of this paper is to provide a Monte Carlo simulation to evaluate the extent to which the SAOM model is able to estimate the level of spatial autocorrelation as commonly assumed in binary spatial autoregressive models.\footnote{The authors wish to acknowledge the DJEI/DES/SFI/HEA Irish Centre for High-End Computing (ICHEC) ({\tt https://www.ichec.ie}) for the provision of computational facilities and support.}
We produce data with spatial autocorrelation using the formulation in Eq. (\ref{eq:bsar}), with different degrees of autocorrelation, ranging from high negative autocorrelation ($\rho=-0.8$), to no autocorrelation ($\rho=0$), to high positive autocorrelation ($\rho=0.8$) and with different sample sizes ($n=50$, $n=250$ and $n=500$). This allows us to evaluate the ability of the estimator to assess the level of autocorrelation and the extent to which this improves as sample size increases. We provide more simulations with values of $\rho$ closer to zero, as most empirical clustering in the social sciences is not of such high magnitude. Similar to \citet{CalabreseElkink2014}, our model includes one independent variable $X \sim N(2,4)$ with an intercept $\beta_1=4$ and a slope parameter $\beta_2=-2$,\footnote{Other related Monte Carlo studies also include one or two independent variables in addition to the spatial effect, such as \citet{DoreianTeuterWang1984}, \citet{BeronVijverberg2004}, \citet{Bille2013}, and \citet{FranzeseHaysCook2016}.} such that the expected value of $\rY^*$, without spatial autocorrelation, is $0$, and we have a balanced binary dependent variable $\rY$.

The spatial (or network) structure in the data, $\mw$, is created following the algorithm used in \citet{BeronVijverberg2004} and \citet{CalabreseElkink2014}, generally known as the random geometric network \citep{DallChristensen2002, Penrose2003}. Here, observations are randomly distributed using a uniform distribution over a square region and $w_{ij}=1$ for all neighbouring locations $j$ that are within a specified distance $d$ from $i$, whereby $d$ is set in such a manner as to keep the average degree of the network at approximately 5.

\begin{figure}
\begin{center}
\includegraphics[width=\textwidth]{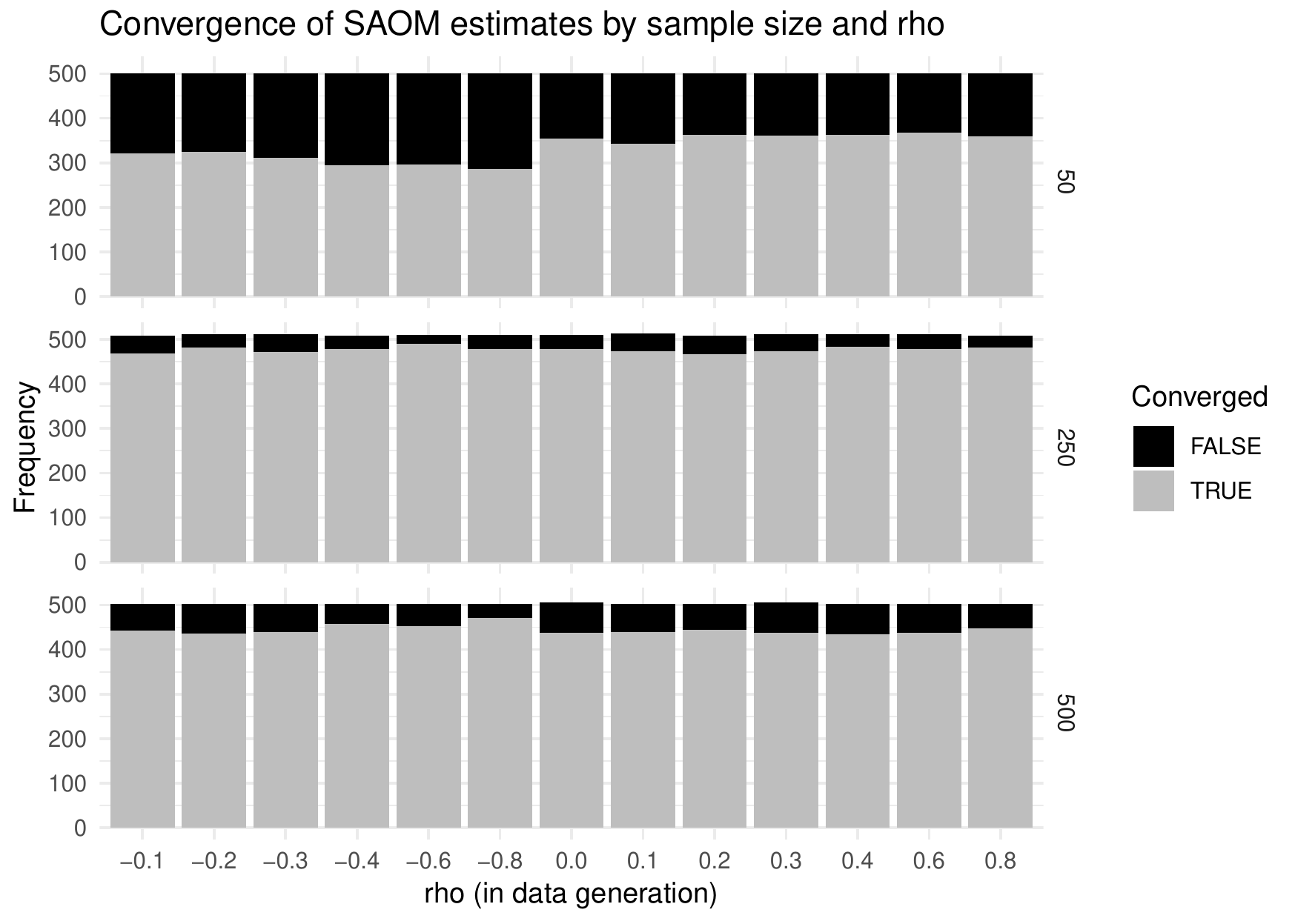}
\caption{Convergence rates for SAOM estimates, by spatial parameter $\rho$ in the data generation process, for different sample sizes.}
\label{fig:convergence}
\end{center}
\end{figure}

In line with common practice for SAOM estimations, simulations where the maximum convergence statistic $t$ is greater than 0.2 or where the convergence statistic for the spatial parameter $\rho$ is greater than 0.1 have been discarded.\footnote{Note that estimations in the Monte Carlo were executed for a maximum of one hour each.} Figure \ref{fig:convergence} provides an overview of convergence rates. The SAOM and the Gibbs samplers  were set to run for 4,000 sampling iterations. On average, SAOM estimates took approximately \timeSAOMs ($\sigma=\timeSAOMsSD$) minutes for $n=50$ and \timeSAOMl ($\sigma=\timeSAOMlSD$) minutes for $n=500$, while Gibbs estimates took \timeGIBBSssec ($\sigma=\timeGIBBSsSDsec$) and \timeGIBBSlsec ($\sigma=\timeGIBBSlSDsec$) seconds, respectively.\footnote{Gibbs estimates are based on the {\tt spatialprobit} package in R \citep{WilhelmGodinhoDeMatos2015}.} 

\begin{figure}
\begin{center}
\includegraphics[width=\textwidth]{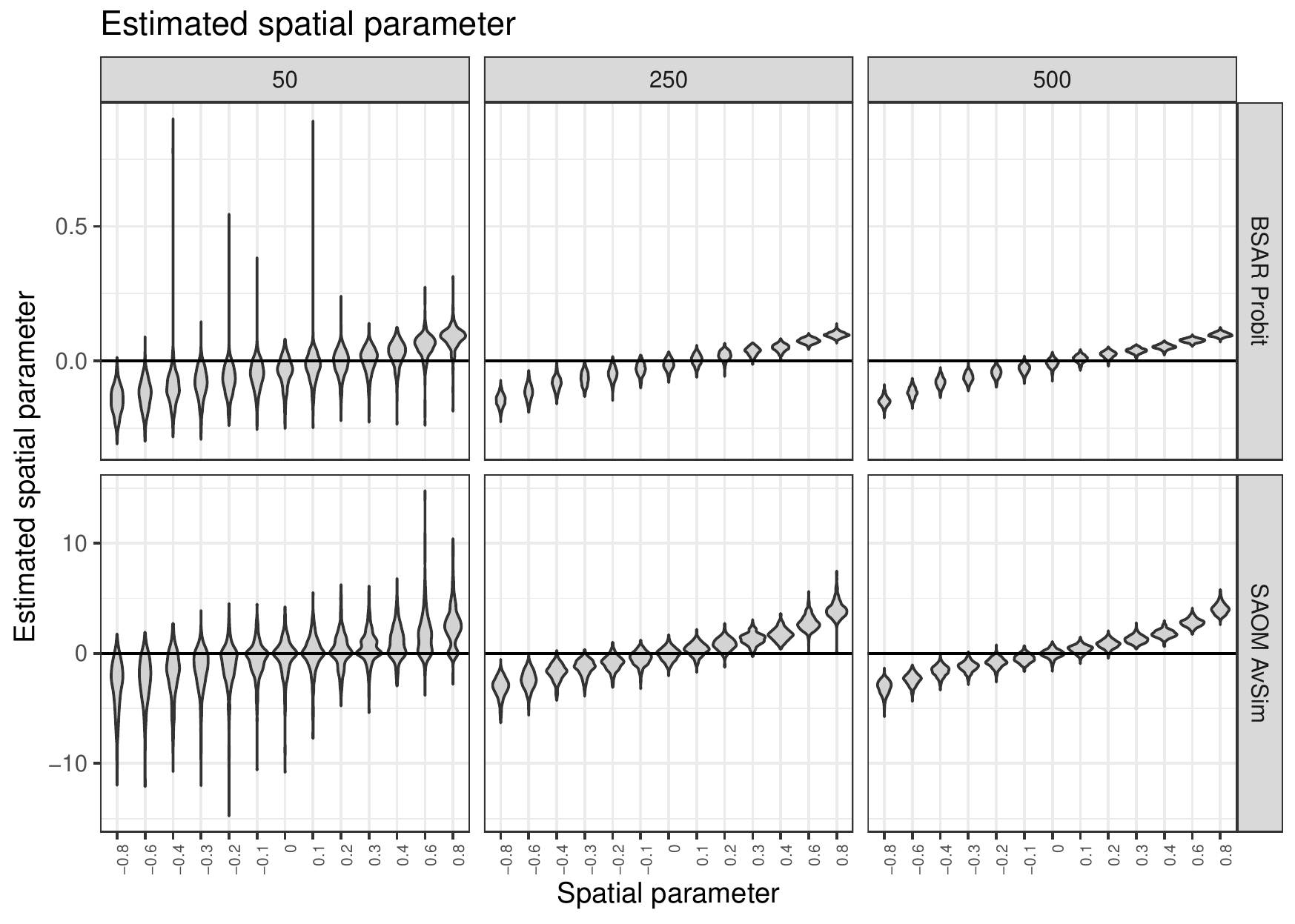}
\caption{Distribution of spatial autoregressive parameter estimates, by spatial parameter $\rho$ in the data generation process, for different sample sizes and for both the BSAR Gibbs and the SAOM model using average similarity. Plus signs mark the value in the data generation process.}
\label{fig:violins}
\end{center}
\end{figure}

As described in the previous section, we estimate the spatial autocorrelation parameter using two different statistics in the SAOM repertoire, namely {\it avSim} and {\it avAlt}. However, since results for {\it avSim} and {\it avAlt} are near-identical, with $\hat{\rho}^{avSim}\approx 2\hat{\rho}^{avAlt}$, we only report on {\it avSim} results. The main estimation results, the quality of the estimation of the intensity of the autocorrelation, are provided in Figure \ref{fig:violins}.\footnote{Table \ref{tab:results} in the online appendix provides means and standard deviations across simulations.} The main result is a very positive one. The estimate of the spatial autoregressive parameter increases with $\rho$, the parameter used in the data generation process, and the estimates improve as the sample size increases, suggesting consistency. Estimates of $\rho$ are indeed highly accurate at a scale $\hat{\rho}^{avSim}\approx 4\rho$. As discussed, one would not expect the spatial autoregressive parameter to be identical to $\rho$ in the data generation process, since in the BSAR specification, the autoregression is in the latent variable $\rY^*$, while in the SAOM specification, nodes react to the actual outcome variable $\rY$ in the neighbouring nodes. \comment{Do we still need a short discussion about how to convert the parameters to the same scale---or how this is not possible? Note that when you take the exponent of $\rho=0.8$ after transformation to logit scale (which I think you cannot really do for an autoregressive parameter), so that we have $e^{0.8\cdot 1.67}=3.80$, we're actually not that far off the {\it avSim} estimate. But $\rho$ should always be between -1 and 1, also when switching from probit to logit, and indeed in \citet{CalabreseElkink2014} we only applied the transformation to the slope and intercept coefficients. Check \citet[167--168]{Ripley_etal2018}.}

\comment{The TNAM results also correlate with $\rho$, but contain a number of very significant outliers, which results in high standard deviations in Table \ref{tab:results} and which is why these results have been excluded from Figure \ref{fig:violins}. The TNAM results are therefore somewhat less reliable as an estimator of the level of spatial autocorrelation, but the estimates are significantly faster.}

\begin{figure}
\begin{center}
\includegraphics[width=.9\textwidth]{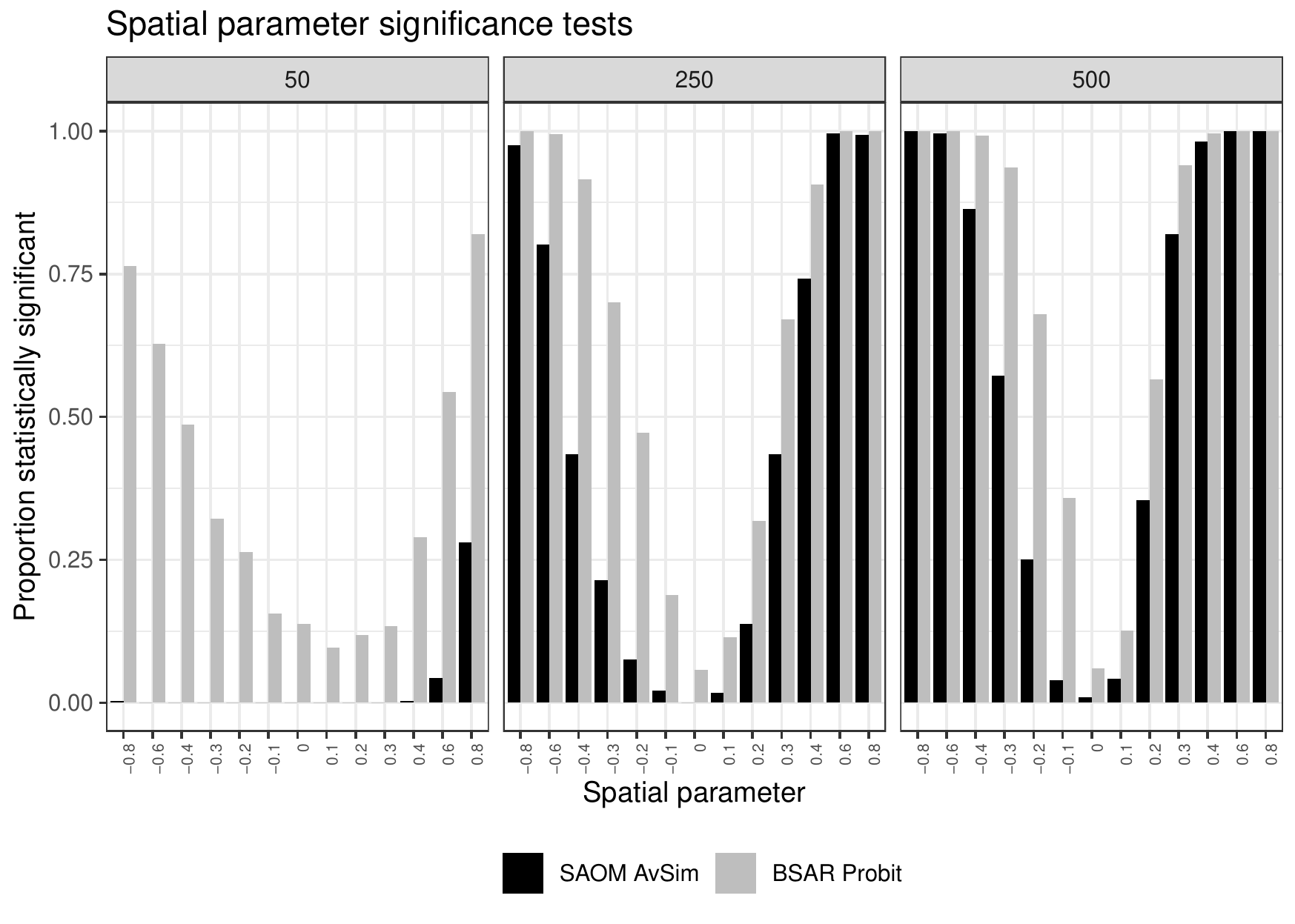}
\caption{Proportion of statistically significant tests on the estimated spatial parameter, for different values of data generation parameter $\rho$, the sample size, and both the BSAR Gibbs and the SAOM models.}
\label{fig:sig}
\end{center}
\end{figure}

While the estimation of the intensity of the autoregression is one concern, the second concern is the ability to detect the autocorrelation in terms of the associated statistical significance test. Ideally, when $\rho=0$ we never detect a significant level of autocorrelation and when $\rho\neq 0$ we always get a statistically significant result---but one would expect the power to detect the autocorrelation to increase with $|\rho|$ and $n$. Figure \ref{fig:sig} provides our results given the estimated spatial autoregressive parameters and the associated estimated standard errors. We see that the statistical significance tests generally fail to detect the spatial autocorrelation when the sample size is small ($n=50$), despite the reasonably accurate estimation on average reported above, but improves as $n$ gets larger ($n=250$ and $n=500$). The false positive rate is low---for BSAR probit it is around the expected 5\% level, while for SAOM it is close to zero. As expected, for low values of $|\rho|$, significance tests often do not detect the autocorrelation. Interestingly, for positive autocorrelation---which is the more common scenario---SAOM considerably outperforms Gibbs in terms of detecting the autocorrelation. Since low, positive values of $\rho$ are the most common in the social sciences, this is a relevant finding. For negative values of $\rho$, Gibbs considerably outperforms SAOM in correctly rejecting the null hypothesis.

\begin{figure}
\begin{center}
\includegraphics[width=\textwidth]{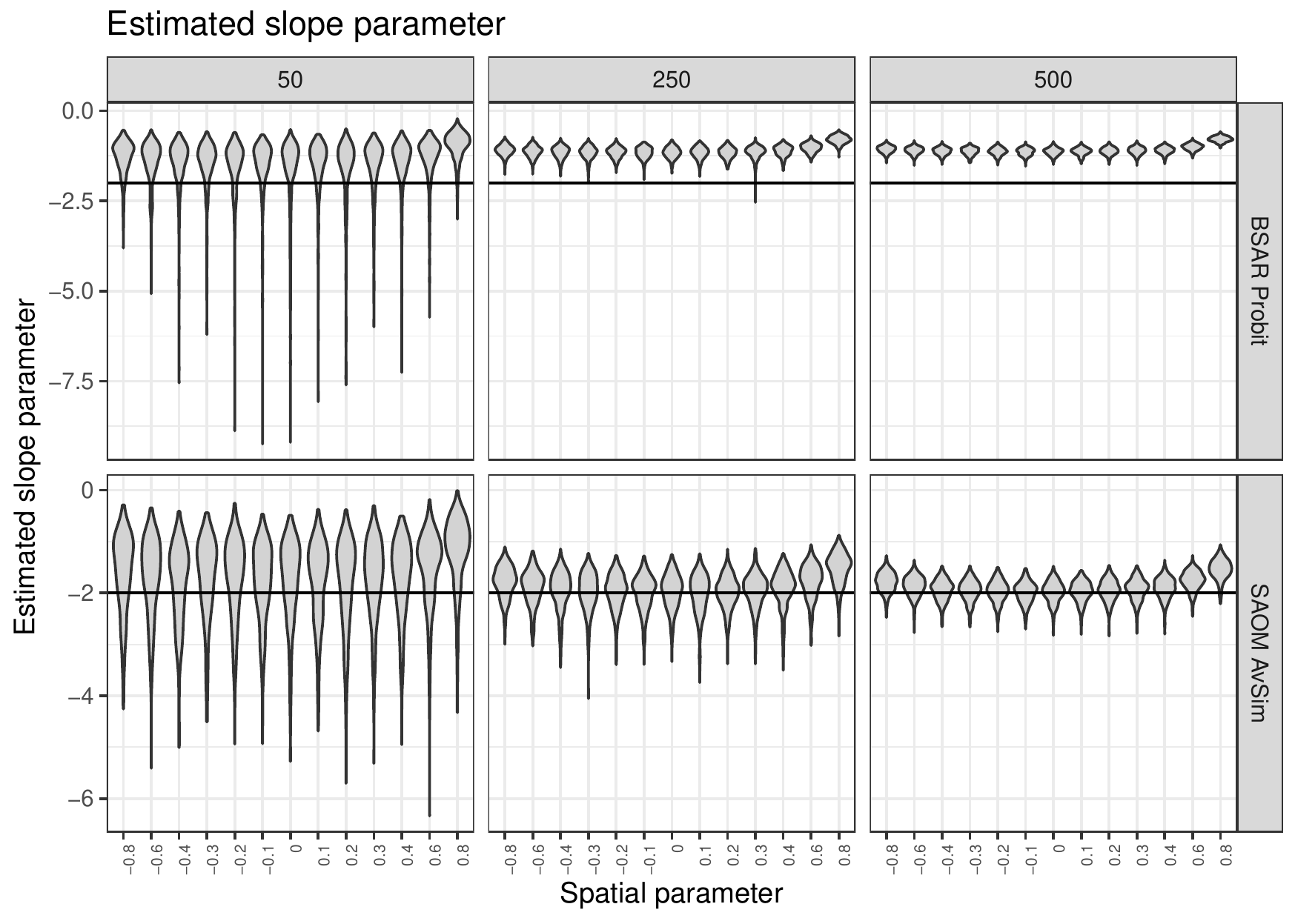}
\caption{Distribution of slope parameter estimates, by spatial parameter $\rho$ in the data generation process, for different sample sizes and statistics used in the SAOM model. Standard deviation in parentheses. The value in the data generating process is -2.} 
\label{fig:slope}
\end{center}
\end{figure}

While the primary focus is on the estimation of the spatial autoregressive parameter, any reasonable application in social science will also require an accurate estimation of the parameter on additional independent variables. When we continue to take the baseline spatial model specification of Eq. (\ref{eq:bsar}) as our baseline, we would at least have node-level exogenous variables in our model and would be required to accurately estimate the slope coefficients for those variables. Figure \ref{fig:slope} provides the analogous estimates for a randomly generated independent variable $\mx$, with a slope parameter in the data generation process of -2. 
Here we find that while the sign of the slope coefficient is correct, estimates tend to be underestimated, except when the sample size is sufficiently large and the spatial autocorrelation not too strong. Remarkably, however, the SAOM estimates are closer to the parameter in the data generation process than the Gibbs estimates, which is found to be one of the best estimators for BSAR specifications \citep{CalabreseElkink2014}. 
Both BSAR and SAOM statistical significance tests on the slope parameter reject the null hypothesis correctly, except when the sample size is particularly small ($n=50$), SAOM often does not have sufficient power to reject the null hypothesis---see Figure \ref{fig:sig_slope} in the online appendix.

Overall we can conclude that the SAOM estimator, in a static, cross-sectional context, performs particularly well in terms of detecting the spatial autocorrelation and correctly estimating the slope coefficients. With low, positive autocorrelation---the most common scenario in social science---it in fact outperforms the Gibbs sampler for spatial econometric models. The cost is the computational complexity and associated slow estimation, but not prohibitively so, and this is compensated by the straightforward parallelisation of SAOM, so that it fully leverages multi-core processors.

\section{Empirical application: The diffusion of same-sex marriage}

\begin{figure}
	\begin{center}
		\includegraphics[width=\textwidth]{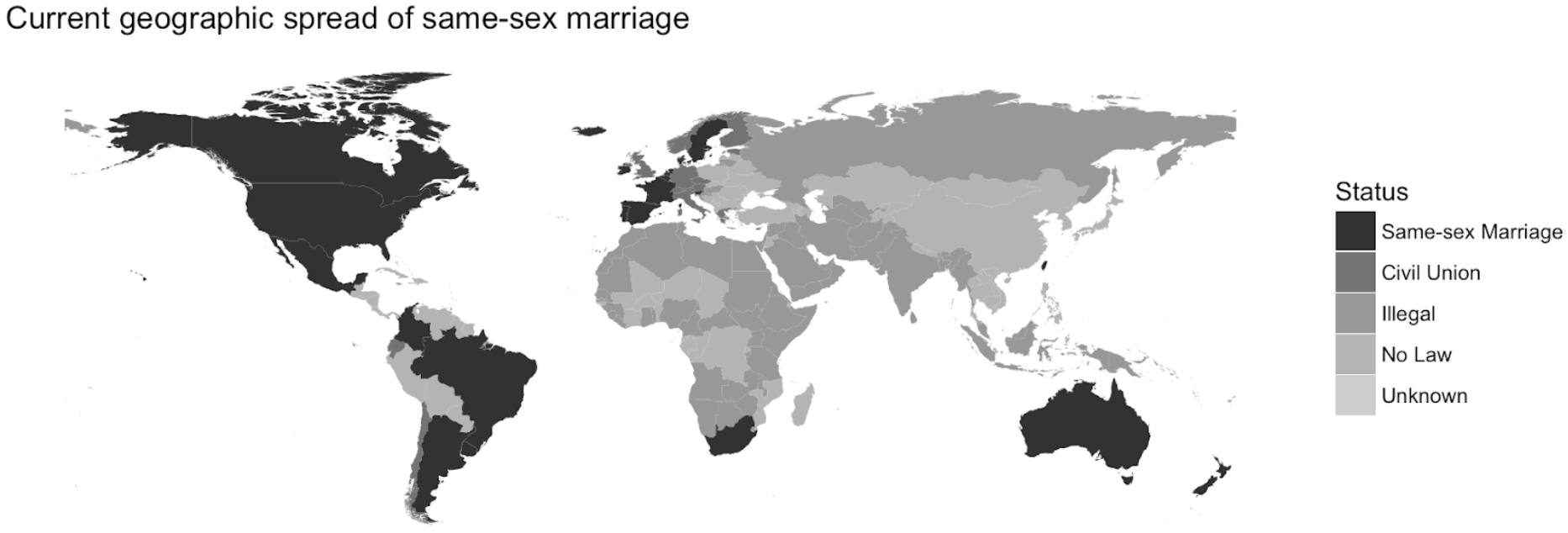}
		\caption{Map of the international spread of civil unions and marriage for same-sex couples.}
		\label{fig:ssm}
	\end{center}
\end{figure}

As we mention in the introduction, a typical application for this type of model is the international diffusion of policies and regimes. Recent years have shown a rapid rise in the availability of civil union and marriage options for same-sex couples, which appears to follow a similar pattern of diffusion \citep[cf][]{HaiderMarkel2001}. Countries take each other's examples and implement liberalizing policies towards same-sex marriage. Typically the diffusion mechanism appears to be through public opinion, where governments are under pressure from a liberalizing population to implement the relevant policies. Populations tend to learn primarily from geographically proximate or culturally similar countries---here we investigate the role of geographic proximity. Figure \ref{fig:ssm} provides an impression of the dependent variable, depicting the current geographic clustering of the availability of same-sex marriage and civil union.

Data about same-sex marriage policy adoption in each nation or state was collected using legislative records, newspapers and other media coverage of judicial or legislative decisions, court briefs and syllabi, international interest group data, and academic journals. On government websites, legislation about same-sex marriage and civil unions were traced to determine dates and votes. In the case of a total lack of information about same-sex marriage or civil union legislation, LGBTQ interest groups listed as members of the International Lesbian, Gay, Bisexual, Trans and Intersex Association were contacted in each nation requesting for the missing information.

The process of the adoption of legislation varies significantly per country. For example, Spanish same-sex marriage went through the legislative processes in less than one year. In France, the law was passed by the legislature, but was reviewed by the higher courts to determine constitutionality, delaying the enactment date. In comparison, Colombia had various court cases that legalized portions of civil unions, with one major ruling enacting the law, inter alia legalizing same-sex marriage as well. This initiated another legal battle to determine the legality of same-sex marriage, preventing the enactment of the law for three more years after the first ruling. Rather uniquely, Ireland voted for the implementation of same-sex marriage by popular referendum \citep{Elkink_etal2017}. Illegality is coded for any nation where there is a written law against homosexuality, same-sex marriage, or any other type of formal punishment for homosexuality. Civil unions encompasses domestic partnerships, same-sex unions, civil unions, and registered partnerships, as all terms are very similar in rights granted per nation.

Many studies have outlined factors that are critical for creating an environment and culture in which same-sex marriage policy diffusion can take place such as public and political ideology, influence of LGBTQ groups, acceptance of homosexuality, religiosity, and international pressure, specifically from the European Union or the European Court on Human Rights \citep{Kollman2007, Kollman2017, Cooper2012, FernandezLutter2013, Kreitzer_etal2014, Ayoub2015, MitchellPetray2016, Hildebrandt_etal2017, SlootmaeckersSircar2018}.

Gay and lesbian issues are considered morality issues due to their conscientious appeal, making their debate and policy diffusion similar to that of other morality policies \citep{HaiderMarkel1999}. Ideas about which demographics support LGBTQ people like women, higher income, and political ideology and determinants such as partisanship and ideology provided the most obvious relationships for lesbian and gay support in traditionally assumed ways \citep{HaiderMarkel1999, GainesGarand2010}. 

Religiosity, or the cultural reinforcement of religious ideas due to religious service attendance, has often been described as a critical factor for determining if same-sex marriage policy can diffuse. Religiosity can slow down the speed at which same-sex marriage or union policies can become law in a nation, as is exemplified by specific types of religions constantly reinforcing ideologies against homosexuality depending on level of church attendance \citep{Kollman2007, HoogheMeeusen2013}. Specific types of religions depicting homosexuality in negative ways repeatedly overtime can create a public disapproval of same-sex marriage, preventing the policy to diffuse from one nation to another \citep{BeerCruzAcevez2018, DionDiez2017}. This could be why highly religious nations like Ireland and the United States took longer to adopt same-sex marriage than low religious nations like the Netherlands \citep{Hildebrandt_etal2017}.

Public tolerance of homosexuality in other countries can also help to determine how same-sex marriage policy diffuses. It can be concluded that a lack of support for gays and lesbians in general will equate to lower levels of support for their rights, including same-sex marriage \citep{GainesGarand2010, BeerCruzAcevez2018}. Specifically, tolerance and support of gays and lesbains demonstrated higher support for same-sex marriage support, indicating that same-sex marriage policies can diffuse in areas that support gays and lesbians rights \citep{GainesGarand2010, HoogheMeeusen2013}. This may help to explain why in nations like the United States, same-sex marriage policy diffused slowly over a variety of states before the policy was adopted by the entire nation, as not all of the states support the policy at the same time.

Our dependent variable is coded as a one for all country-years where the country has either the possibility of civil union or of marriage for same-sex couples and zero in all other cases. We control for a number of socio-economic variables and a number of variables that can be expected to correlate with liberal values, including religion. The first include population size,\footnote{Sourced from the Pew Research Center for all world religions in 2010.} employment and life expectancy.\footnote{Both sourced from the World Development Indicators of the World Bank.} To proxy for general liberal values we include the ratio of female to male labour participation and proportion of seats held by women in national parliaments,\footnote{Both sourced from the World Development Indicators of the World Bank.} as well as the relative percentage of Christians and Muslims.\footnote{Sourced frmo the Pew Research Center data.} For all independent variables we apply linear interpolation, over time within country, to fill in missing data, while population and religion data is fixed over the full twenty years.

\begin{table}
	\begin{center}
\begin{tabular}{l|r|rr|rr|rr}
	& & \multicolumn{4}{c|}{Static (2018)} & \multicolumn{2}{c}{Dynamic} \\
	\hline
	& Probit & \multicolumn{2}{c|}{BSAR Gibbs} & \multicolumn{4}{c}{SAOM avSim}  \\
	\hline
	Geographic proximity & & 0.09** & 0.03* & 3.11** & 3.34** & 3.27** & 2.59** \\
	& & (0.01) & (0.02) & (0.20) & (0.83) & (0.24) & (0.32) \\
	Log of population & 0.05 & & 0.10 & & 0.32* & & 0.15** \\
	& (0.07) & & (0.08) & & (0.19) & & (0.07) \\
	Employment & -0.03 & & -0.03* & & -0.01 & & -0.03* \\
	& (0.02) & & (0.02) & & (0.05) & & (0.02) \\
	Life expectancy & 0.11** & & 0.10** & & 0.06 & & 0.13**\\
	& (0.03) & & (0.02) & & (0.07) & & (0.03) \\
	Female labour force & 0.03* & & 0.03* & & 0.03 & & 0.05** \\
	& (0.01) & & (0.02)& & (0.03) & & (0.01) \\
	Female seats in parl. & 0.02 & & 0.01 & & 0.02 & & 0.06** \\
	& (0.01) & & (0.01) & & (0.03) & & (0.01) \\
	Proportion Christian & 0.01 & & 0.01 & & -0.00 & & 0.00 \\
	& (0.01) & & (0.01) & & (0.02) & & (0.01) \\
	Proportion Muslim & -0.02 & & -0.01  & & -0.00 & & -0.04** \\
	& (0.02) & & (0.01) & & (0.02) & & (0.01) \\
	{\it Intercept} & -11.56** & -0.34** & -10.90** & & & & \\
	& (2.83) & (0.10) & (2.56) & & & & \\
	{\it Linear shape} & & & & & & -0.27** & -2.58** \\
	& & & & & & (0.12) & (0.37) \\
	\hline
	$N$ & 188 & 188 & 188 & 188 & 188 & 188 & 188 \\
	$T$ & 1 & 1 & 1 & 1 & 1 & 20 & 20 \\
\end{tabular}
\caption{Empirical analysis of the diffusion of the legalization of civil union and marriage for same-sex couples.}
\label{tab:empirics}
\end{center}
\end{table}

We first investigate the empirical data in a static manner, along the lines of the above Monte Carlo simulations. Estimating both BSAR and SAOM models using data on the presence of civil unions and marriage for same-sex couples in 2018, we indeed find evidence of a geographic clustering.
Table \ref{tab:empirics} provides our regression results, which show a clear pattern of geographic diffusion of civil unions and same-sex marriage.\footnote{By default, the SAOM estimator imputes missing values on covariates by the overall mean (\citealt[33--4]{Ripley_etal2018}; see also \citealt{HuismanSteglich2008}). For reasons of comparability, we apply the same strategy for the probit estimates.} The results are particularly interesting keeping in mind the various simulation results. Our Monte Carlo analysis shows that SAOM is better equiped to detect statistically significant levels of clustering when the spatial clustering is indeed present in the data generation process, but is of low magnitude. The results in Table \ref{tab:empirics} suggest this is indeed the case. The SAOM estimator also had lower power on significance tests for the slope parameters when the sample size is low, which is the case here. The signs of the coefficients on the explanatory variables are largely identical to those in the probit regression, but with larger standard errors.

The regression analysis therefore suggests that spatial clustering is present in the spread of same-sex marriage and civil unions, but is sufficiently weak not to be detected by the Gibbs sampler. The key advantage of turning from binary spatial autoregressive models to the stochastic actor-oriented model is the straightforward ability to implement temporal dynamics. We therefore continue by estimating the SAOM model for the full 1999--2018 period. Because of the low number of changes in the dependent variable in each year, we apply a similar strategy to the static estimate of the SAOM estimator. While this time we do not artificially alter any values in the dependent variable, we do again fix the rate of change in behaviour at one for all years and the rate of network change at an arbitrarily small value. This ensures that even with few changes in the dependent variable, sufficient opportunities for change occur, allowing the estimator to estimate the parameters that govern this change---or lack thereof. The results are striking. With a proper dynamic model, we continue to find clear evidence of spatial clustering, but also statistically significant results on nearly all control variables, in the expected direction.

\section{Concluding remarks}

Overall simulations results suggest that statistical network models, that are originally designed for dynamic analysis with as key endogenous variable the network ties, can be used in a static context to estimate the level of autocorrelation in node characteristics. A slight tweak to the SAOM model specification allows one to estimate this model and results are in line with the data generation process. Furthermore, in particular as the sample size increases to moderate size, statistical significance tests correctly identify the spatial autocorrelation as long as the strength is reasonably strong.

There are a number of important reasons to prefer statistical network models over spatial econometric ones. Most importantly, the fact that the statistical network model incorporates a reaction to the actual value of the binary dependent variable in neighbouring nodes, as opposed to the latent variable, is a much more natural and theoretically appropriate specification. Secondly, the extension to a dynamic context is significantly more straightforward than with spatial regression models. While \citet{FranzeseHaysCook2016} propose an estimator for spatial context, using a specialised $\mw$ matrix to incorporate temporal effects akin to a lagged dependent variable in regular time-series analysis, the SAOM estimator is designed for dynamic modelling. A further benefit is that the use of a statistical network model implies that all complications from statistical network modelling can easily be incorporated, including co-evolution and high order interdependences between network structure and node characteristics. In particular when the spatial econometric model is applied to networks other than geographical contiguity, these factors become highly relevant. When the network is not distance but trade ties, or cultural similarity, or development aid flows, or migration flows, these network ties significantly vary over time, sometimes for reasons closely correlated with factors that influence the dependent variable, whether it is the political regime or particular policies.

\bibliographystyle{apsr}
\bibliography{sl-network}

\newpage
\appendix

\section{Online appendix}

\begin{figure}
\begin{center}
\includegraphics[width=.9\textwidth]{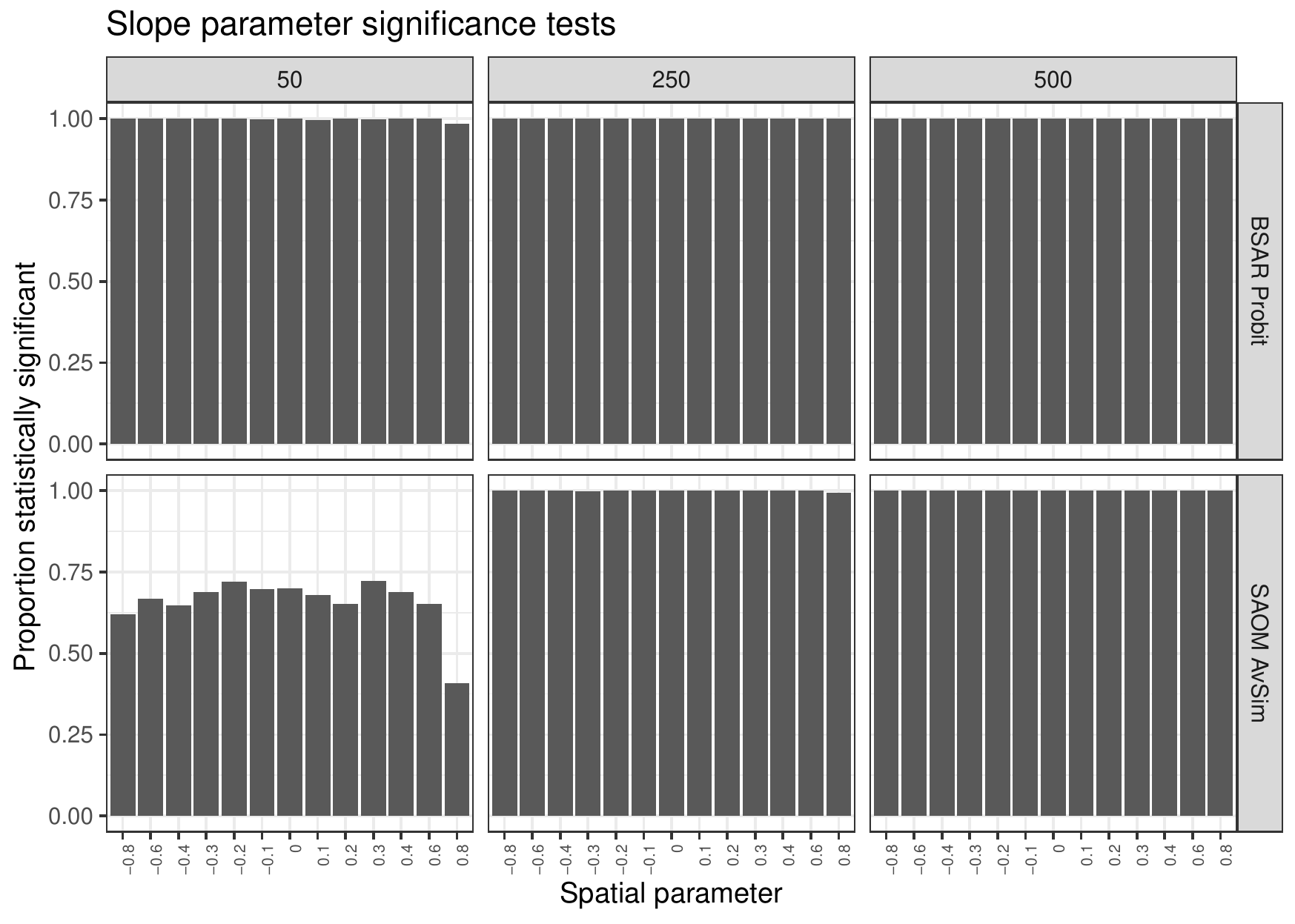}
\caption{Proportion of statistically significant tests on the slope parameter on $\mx$, for different values of data generation parameter $\rho$, the sample size, and both the BSAR Gibbs and the SAOM models.}
\label{fig:sig_slope}
\end{center}
\end{figure}

This online appendix provides a number of additional figures and tables related to our Monte Carlo analysis. Table \ref{tab:counts} provides the overall number of simulation results, after simulations that did not converge have been removed. This provides the same information as Figure \ref{fig:convergence} in the paper. Table \ref{tab:results} provides the mean and standard deviation across replications for all estimates of the autocorrelation parameter $\rho$, which is a numerical summary of the distributions presented in Figure \ref{fig:violins} in the paper. Table \ref{tab:results_slope} is the equivalent table for the slope parameter on the exogenous explanatory variable $\mx$. Figure \ref{fig:sig_slope} provides the proportion of statistically significant tests on the slope parameter, where the true value in the data generation process is -2. We find that for all but the smallest samples, both estimators correctly reject the null hypothesis.

\begin{table}[!h]
\begin{center}
\begin{tabular}{c|c|c|c|c|c|c|}
&\multicolumn{3}{c|}{BSAR Probit}&\multicolumn{3}{c|}{SAOM AvSim}\\
&50&250&500&50&250&500\\
\hline
-0.80&500&500&500&287&475&463\\
-0.60&500&500&500&296&479&443\\
-0.40&500&500&500&294&469&453\\
-0.30&500&500&500&311&463&428\\
-0.20&500&500&500&324&475&431\\
-0.10&500&500&500&321&463&433\\
0.00&500&500&500&354&468&430\\
0.10&500&500&500&343&466&431\\
0.20&500&500&500&362&459&435\\
0.30&500&500&500&361&465&427\\
0.40&500&500&500&362&473&429\\
0.60&500&500&500&367&467&433\\
0.80&500&500&500&360&468&440\\
\end{tabular}

\caption{Total number of simulation results, by spatial parameter $\rho$ in the data generation process, for different sample sizes and estimators. The lower number of SAOM estimates is due to the omission of results that did not converge.}
\label{tab:counts}
\end{center}
\end{table}

\begin{table}
\begin{center}
\begin{tabular}{c|c|c|c|c|c|c|}
&\multicolumn{3}{c|}{BSAR Probit}&\multicolumn{3}{c|}{SAOM AvSim}\\
&50&250&500&50&250&500\\
\hline
-0.80&-0.15&-0.15&-0.15&-2.86&-3.10&-3.10\\
&(0.05)&(0.02)&(0.02)&(2.27)&(0.91)&(0.62)\\
-0.60&-0.13&-0.11&-0.12&-2.42&-2.40&-2.37\\
&(0.06)&(0.03)&(0.02)&(2.16)&(0.84)&(0.52)\\
-0.40&-0.10&-0.08&-0.08&-1.82&-1.69&-1.60\\
&(0.09)&(0.03)&(0.02)&(2.11)&(0.74)&(0.50)\\
-0.30&-0.08&-0.06&-0.06&-1.47&-1.28&-1.22\\
&(0.06)&(0.03)&(0.02)&(1.92)&(0.71)&(0.45)\\
-0.20&-0.07&-0.05&-0.04&-1.08&-0.89&-0.84\\
&(0.06)&(0.03)&(0.02)&(2.03)&(0.67)&(0.45)\\
-0.10&-0.05&-0.03&-0.03&-0.54&-0.46&-0.47\\
&(0.06)&(0.02)&(0.02)&(1.57)&(0.63)&(0.40)\\
0.00&-0.04&-0.01&-0.01&-0.30&-0.08&-0.03\\
&(0.05)&(0.02)&(0.02)&(1.64)&(0.58)&(0.39)\\
0.10&-0.02&0.00&0.01&-0.07&0.36&0.38\\
&(0.07)&(0.02)&(0.01)&(1.65)&(0.56)&(0.36)\\
0.20&-0.01&0.02&0.02&0.26&0.83&0.83\\
&(0.05)&(0.02)&(0.01)&(1.54)&(0.57)&(0.38)\\
0.30&0.00&0.04&0.04&0.61&1.29&1.27\\
&(0.05)&(0.01)&(0.01)&(1.42)&(0.56)&(0.35)\\
0.40&0.03&0.05&0.05&0.99&1.73&1.78\\
&(0.04)&(0.01)&(0.01)&(1.47)&(0.57)&(0.36)\\
0.60&0.06&0.07&0.08&1.65&2.78&2.79\\
&(0.04)&(0.01)&(0.01)&(1.86)&(0.71)&(0.38)\\
0.80&0.09&0.10&0.10&2.55&3.99&4.05\\
&(0.04)&(0.01)&(0.01)&(1.90)&(0.84)&(0.50)\\
\end{tabular}

\caption{Distribution of spatial autoregressive parameter estimates, by spatial parameter $\rho$ in the data generation process, for different sample sizes and estimators. Standard deviation in parentheses.}
\label{tab:results}
\end{center}
\end{table}

\begin{table}
\begin{center}
\begin{tabular}{c|c|c|c|c|c|c|}
&\multicolumn{3}{c|}{BSAR Probit}&\multicolumn{3}{c|}{SAOM AvSim}\\
&50&250&500&50&250&500\\
\hline
-0.80&-1.30&-1.10&-1.07&-1.62&-1.79&-1.79\\
&(0.48)&(0.15)&(0.10)&(0.77)&(0.31)&(0.19)\\
-0.60&-1.43&-1.14&-1.11&-1.69&-1.90&-1.87\\
&(0.61)&(0.16)&(0.10)&(0.77)&(0.34)&(0.19)\\
-0.40&-1.49&-1.16&-1.13&-1.84&-1.95&-1.95\\
&(0.78)&(0.17)&(0.10)&(0.80)&(0.38)&(0.21)\\
-0.30&-1.47&-1.17&-1.13&-1.74&-1.98&-1.96\\
&(0.65)&(0.18)&(0.11)&(0.82)&(0.38)&(0.21)\\
-0.20&-1.57&-1.16&-1.13&-1.73&-1.98&-1.97\\
&(0.81)&(0.15)&(0.11)&(0.79)&(0.35)&(0.21)\\
-0.10&-1.53&-1.17&-1.14&-1.80&-2.00&-1.99\\
&(0.82)&(0.16)&(0.11)&(0.76)&(0.35)&(0.21)\\
0.00&-1.64&-1.16&-1.14&-1.78&-1.97&-1.98\\
&(1.01)&(0.16)&(0.10)&(0.82)&(0.34)&(0.21)\\
0.10&-1.58&-1.17&-1.13&-1.80&-2.02&-1.97\\
&(0.92)&(0.17)&(0.10)&(0.80)&(0.39)&(0.20)\\
0.20&-1.62&-1.17&-1.12&-1.84&-2.00&-1.95\\
&(0.99)&(0.16)&(0.11)&(0.89)&(0.33)&(0.21)\\
0.30&-1.53&-1.15&-1.11&-1.73&-1.96&-1.94\\
&(0.77)&(0.17)&(0.11)&(0.77)&(0.33)&(0.21)\\
0.40&-1.49&-1.12&-1.10&-1.72&-1.91&-1.90\\
&(0.83)&(0.15)&(0.10)&(0.76)&(0.36)&(0.20)\\
0.60&-1.32&-1.03&-1.01&-1.55&-1.78&-1.76\\
&(0.64)&(0.13)&(0.09)&(0.76)&(0.33)&(0.19)\\
0.80&-0.96&-0.80&-0.79&-1.19&-1.53&-1.54\\
&(0.38)&(0.12)&(0.08)&(0.69)&(0.31)&(0.20)\\
\end{tabular}

\caption{Distribution of slope autoregressive parameter estimates, by spatial parameter $\rho$ in the data generation process, for different sample sizes and estimators. The value in the data generating process is -2.} 
\label{tab:results_slope}
\end{center}
\end{table}

\end{document}